\documentclass[aps,pre,twocolumn,showpacs,showkeys,a4paper]{revtex4-1}
 \usepackage{graphicx}
 \usepackage{amsmath}
 \usepackage{amssymb}
 \usepackage{amscd}
 \usepackage{color}

\begin{document} 

 \title{On the motion of kinesin in a viscoelastic medium}
 \author{Gert Knoops$^{(1)}$ and Carlo Vanderzande$^{(1,2)}$}
 \affiliation{(1) Faculty of Sciences, Hasselt University, 3590 Diepenbeek, Belgium\\
  (2) Instituut Theoretische Fysica, Katholieke Universiteit Leuven, 3001
  Heverlee, Belgium}


 \begin{abstract} 
Kinesin is a molecular motor that transports cargo along microtubules. The results of many {\it in vitro} experiments on kinesin-1 are described by kinetic models \cite{Clancy11} in which one transition corresponds to the forward motion and subsequent binding of the tethered motor head. We argue that in a viscoelastic medium like the cytosol of a cell this step is not Markov and has to be described by a non-exponential waiting time distribution. We introduce a semi-Markov kinetic model for kinesin that takes this effect into account. We calculate, for arbitrary waiting time distributions, the moment generating function of the number of steps made, and determine from this the average velocity and the diffusion constant of the motor. We illustrate our results for the case of a waiting time distribution that is Weibull. We find that for realistic parameter values, viscoelasticity decreases the velocity and the diffusion constant, but increases the randomness (or Fano-factor). 
 \end{abstract}

\maketitle 
\section{Introduction}
Molecular motors are proteins that play an important role in various biological processes like cell motion, cell division and intracellular transport of organelles or other cargos \cite{Howard01}.  These motor proteins convert chemical energy into work through the hydrolysis of a nucleotide. In this paper we focus on kinesin-1, which is an ATP-driven motor that makes 
steps on the microtubules of the cytoskeleton. Kinesin is a two-headed protein whose heads are connected with a linker that in turn is connected to a cargo-binding tail. Through numerous {\it in vitro}  experiments, the mechanochemical details of kinesin stepping have been well characterised \cite{Svoboda93,Asbury03,Kaseda03,Yildiz04,Block90}. These studies have led to a standard discrete state Markov model for the motion of kinesin-1 \cite{Clancy11}. In the simplest version of this model, the dynamics of kinesin-1 can be described in terms of three kinetic states. Transitions between these states are given by experimentally determined rates. An important role in kinesin's motion is played by the mechanical properties of the neck linker. It is thought that after the binding of ATP to the front head, the free head has to perform a diffusive motion to the next site on the microtubule where it can then bind. Since the linker has to stretch to reach this site, this diffusive motion depends on its elastic properties. In the simplest approximation, the linker is often modelled as a Hookean spring. More realistically, it can be described in terms of a wormlike chain \cite{Kutys10}. A good model for the motion of the tethered head and its subsequent binding is therefore that of a particle moving in a double well potential (see section 2 for more details). The associated rate in the kinetic model is then given in terms of Kramers' transition rate \cite{Kramers40,Hanggi91,Keller00}.

The motion of motors inside real cells have been investigated less \cite{Cai07,Hill04}. In a recent {\it Perspective} in The Biophysical Journal, the author asks how it is possible that the speed of kinesin-1 motors is hardly influenced by the crowded environment of the cell \cite{Ross16}. Indeed, it has by now been well established that cytoplasmic crowdedness leads to viscoelasticy \cite{Weiss13} which in turn makes the passive motion of various proteins, organelles, etc. in the cell subdiffusive \cite{Hofling13}. In this paper we want therefore to investigate if and how viscoelasticity influences the active motion of kinesin-1 along the microtubule. Some {\it in vitro} studies of this issue have been performed in recent years \cite{Gagliano10,Sozanski15}.

Going back to the standard model of kinesin-1, we notice the following. When the diffusive motion of the free head takes place in a viscoelastic environment one has to take into account that motion in such a medium is non-Markovian and is usually described in terms of a memory-dependent friction \cite{Goychuk12}. Memory effects also modify the survival probability inside a potential well. In general, it turns out that the survival time and the related escape rate are no longer exponentially distributed. 
For example, simulations have established that the survival probability of a particle moving in a double well potential  in a viscoelastic environment is very well described by a stretched exponential \cite{Goychuk09}. Hence, a proper way to take into account the effects of crowdedness of the cellular environment in the standard model of kinesin-1 is by replacing the exponential waiting time in the step describing the diffusion of the tethered motor domain with a non-exponential waiting time. 

Discrete state models with these types of waiting times are known as semi-Markov models in the mathematical literature. It turns out that the resulting semi-Markov version of the three-state model of kinesin is still exactly solvable. Moments of the position of the motor can be found
using an approach in which the full moment generating function of the number of steps made by the motor is determined. This technique is well known in the large deviation approach to non equilibrium statistical mechanics \cite{Lebowitz99,Barato15} and works both in the Markov and semi-Markov case. 
Indeed, the standard approach to calculate the diffusion constant \cite{Fisher99,Fisher99b,Kolomeisky00} cannot be easily extended to the semi-Markov case.

We are not the first to investigate the motion of a motor in a viscoelastic environment. In recent work the same problem was studied within the context of ratchet models of molecular motors \cite{Goychuk14a,Goychuk14b}. Indeed, the two well established theoretical approaches to molecular motors are ratchet potential models and discrete-state stochastic models \cite{Kolomeisky13}. While both approaches have their advantages, models based on ratchet potentials can seldomly be solved exactly. Moreover it is hard to determine realistic potentials from experimental data. Finally, it is difficult to include complicated biochemical pathways in these type of models. For these reasons, it is necessary to study the effects of viscoelasticity also within discrete state stochastic models.

This paper is organised as follows. In section 2 we introduce our three-state semi-Markov model of kinesin-1. In section 3 we calculate the average velocity of the motor for a general waiting-time distribution. We find that, as a function of ATP-concentration, the velocity always has the Michaelis-Menten form. We also give specific results for a waiting time distribution that is a Weibull distribution. In section 4 we explain our approach to calculate the moment generating function of the position of the motor. In section 5 we show that quite generally the second cumulant of the position of the motor diffuses. Our model does not have any sub- or superdiffusive motion. For the case of a Weibull waiting time distribution, we calculate the diffusion constant and the Fano factor $F=2D/Vd_0$ (where $d_0$ is the size of the step made, i.e. $4.1$ nm) as a function of the properties of the viscoelastic environment and as a function of ATP-concentration. Finally, in section 6 we present our conclusions.  

\section{The model}
\subsection{Kinetic states}
\begin{figure}[h]
\centering
\includegraphics[width=8cm]{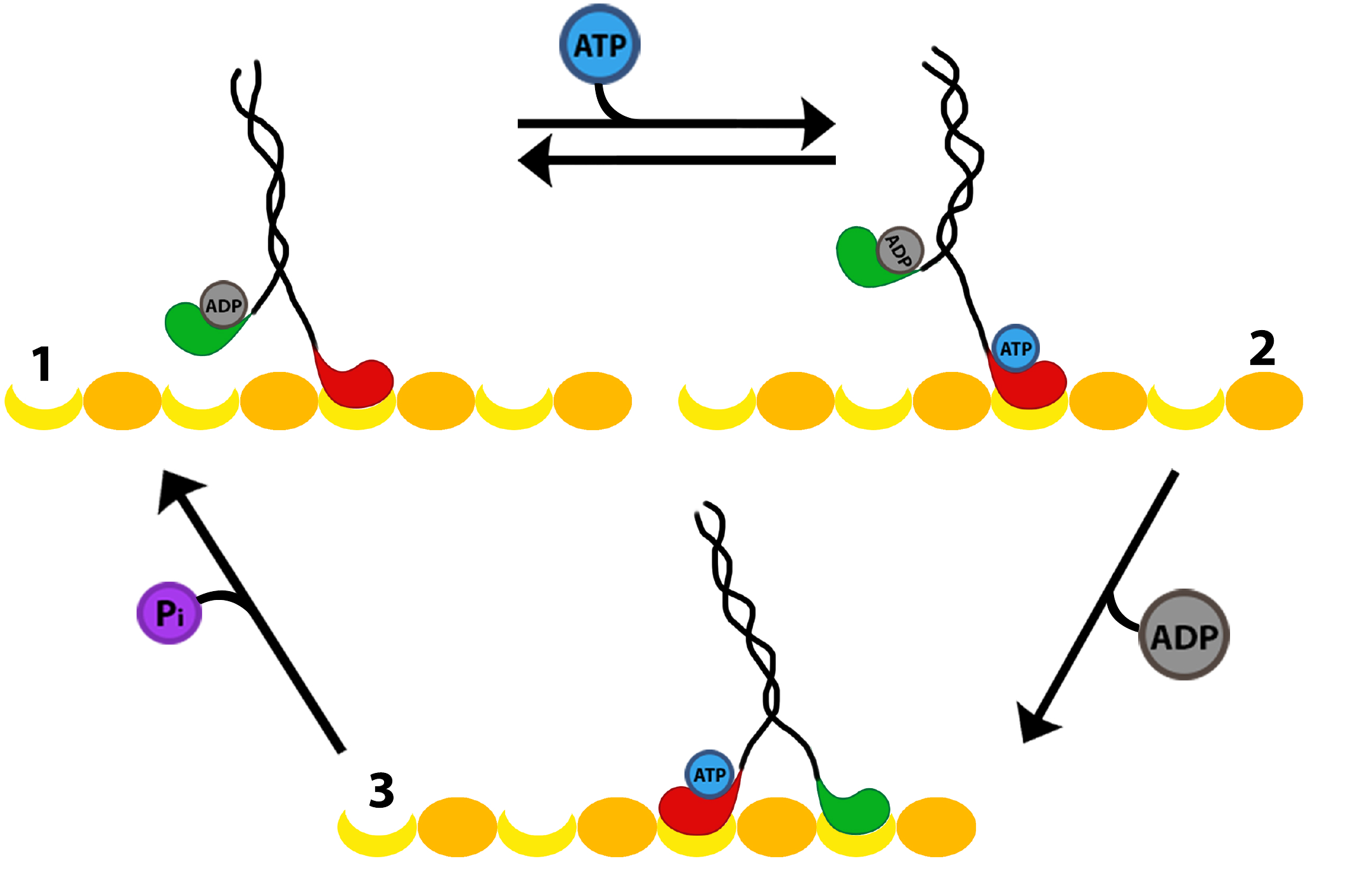}
\caption{Standard three state stochastic model of kinesin-stepping (see text for full explanation).}
\label{Fig1}
\end{figure}

Our starting point is a standard model for the motion of kinesin-1 \cite{Clancy11} which is represented schematically in Fig.\ \ref{Fig1}. In state 1, one motorhead is bound to the microtubule while the second  head is unbound. The binding of ATP to the bound head leads to a conformational change in the motor which allows the free tethered head to move forward (state 2). The transition between states 1 and 2 is dependent on the ATP-concentration and is reversible. As long as ATP is bound to the leading motorhead, the tethered head can move forward and bind to the microtubule (state 3) while releasing an ADP-molecule. This step is determined by the elastic properties of the head linker, by the free energy gained upon binding and by the rheological properties of the environment. 
In the final step ($3\to1$), ATP is hydrolyzed and what is now the trailing head detaches form the microtubule.

If we denote by $M_i$ the conformation of the motor in state $i\ (\in \{1,2,3\})$ and by $T$ and $D$ ATP respectively ADP, we have the following reaction scheme
\begin{eqnarray*}
M_1 +T \rightleftharpoons M_2 \rightarrow M_3 + D \rightarrow M_1
\end{eqnarray*}

\subsection{From viscoelastic environments to non-exponential waiting times}
In this subsection we explain in more detail why the motion of a motor in a viscoelastic environment like the cytoplasm of a cell leads to a description in terms of non-exponential waiting time distributions and semi-Markov processes. 

\begin{figure}[h]
\centering
\includegraphics[width=8cm]{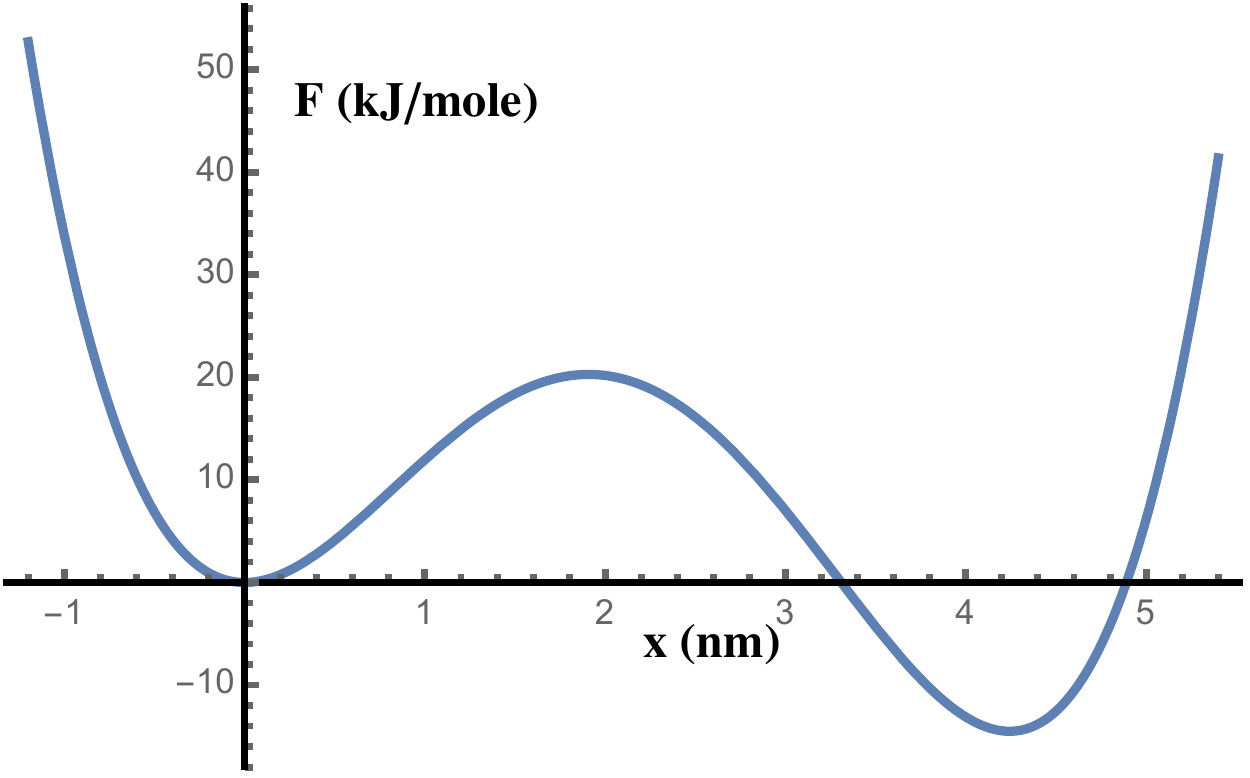}
\caption{Schematic free energy landscape as a function of the position of the motorhead during the transition $2 \to 3$ (fig.1). The left well corresponds to the state $M_2$, the right one to $M_3$.}
\label{Fig2}
\end{figure}

In the transition between the states 2 and 3, the tethered motorhead needs to stretch to reach the binding site. The elastic energy of the linker will therefore increase. When the head binds this leads to a lowering of the free energy. It is therefore reasonable to see the motion of the tethered head as occuring in a double well potential (see Fig. \ref{Fig2}) where the initial increase of the free energy landscape $F(x)$ is due to the stretching of the linker and the second well corresponds to the bound state \cite{Keller00}. 

In a viscoelastic environment the equation of motion for the position $x(t)$ of the tethered head is given by a generalised Langevin equation 
\begin{eqnarray}
m \frac{d^2 x}{dt^2}= -\int_{-\infty}^t K(t-t') \frac{dx}{dt}(t') dt'  - \frac{dF}{dx} + \xi(t)
\label{genlan}
\end{eqnarray}
The first term on the right-hand side corresponds to the viscoelastic friction. In the case that $K(t)=2 \gamma \delta(t)$ one recovers the usual viscous friction term $\gamma dx/dt$. Here $\gamma$ is the friction coefficient which for a sphere of radius $a$ moving through a medium of viscosity $\eta$ is given by Stokes' law
\begin{eqnarray}
\gamma = 6 \pi a \eta
\label{Stokes}
\end{eqnarray}
A good model for a complex viscoelastic medium is to take the kernel $K(t)$ of the form \cite{Goychuk12}
\begin{eqnarray}
K(t) = \gamma (2-\alpha) (1-\alpha) t^{-\alpha}
\end{eqnarray}
Here $\alpha$ is a parameter that characterises the viscoelastic environment. For $\alpha=0$, one has a purely elastic medium, while for $\alpha = 1$ we recover the viscous case. Intermediate values, $0< \alpha < 1$, correspond to the viscoelastic situation. Estimates of $\alpha$ for cells range from $\alpha \approx 0.7$ in {\it E. Coli} \cite{Weber10} to $\alpha \approx 0.2$ in the cytoplasm of eukaryotes \cite{Fakhri14}. 
Finally, the thermal force $\xi(t)$  in (\ref{genlan}) is a Gaussian random variable  with average zero and a correlation that is coupled to the kernel $K(t)$ by the fluctuation-dissipation theorem
\begin{eqnarray}
\langle \xi(t) \xi(t') \rangle = k_B T K(|t-t'|)
\label{fdt}
\end{eqnarray}
where $k_B$ is Boltzmann's constant and $T$ is temperature. 

We now turn to the concept of survival probability and how it depends on the rheological properties of the environment. Image  that at $t=0$ one starts with a large number of particles (representing motor heads) near the bottom of the left well in Fig. \ref{Fig2}. Due to thermal agitation some particles will escape out of that well and move to the right well. Suppose that in the right well the particles are trapped and therefore cannot move back to the left. The probability that a particle is still in the left well at time $t$ is then called the survival probability $S(t)$. With the survival probability, one can associate a waiting time density $\psi_S(t)$ which is such that $\psi_S(t) dt$ gives the probability that a particle escapes from the well in the infinitesimal time interval between $t$ and $t+dt$. Obviously, one has
\begin{eqnarray}
\psi_S(t) = - \frac{d}{dt} S(t)
\label{ws}
\end{eqnarray}
In the viscous case it is well known that asymptotically in time, $S(t)$ is exponential, i.e. $S(t)=\exp(-k t)$ (hence $\psi_S(t)=k \exp(-k t)$) where the escape rate $k$ is given by Kramers' famous formula \cite{Kramers40,Hanggi91} in terms of the viscosity of the environment and properties of the potential such as the height of the barrier between the two minima.

For viscoelastic environments much less is known about the behaviour of $S(t)$ (see the discussion in \cite{Goychuk09}). However,  numerical work \cite{Goychuk09} has shown that in that case the survival probability can be well approximated by a stretched exponential, i.e. one has
\begin{eqnarray}
S(t) = e^{-(k_S t)^\beta}
\label{se}
\end{eqnarray}
where the parameter $k_S$ and $\beta$ depend on the properties of the viscoelastic environment (the exponent $\alpha$ and the friction $\gamma$) and on the height of the potential barrier. Notice that for $\beta=1$ we recover the exponential survival time.
Using (\ref{ws}) we find for the waiting time distribution in this case
\begin{eqnarray}
\psi_S(t)=\beta k_s^\beta t^{\beta-1} e^{-(k_S t)^\beta}
\label{Weibull}
\end{eqnarray}
This distribution is known as the Weibull distribution.

The conclusion of this subsection is therefore: if the motorhead moves through a viscoelastic environment and if we describe that motion in terms of transitions between specific kinetic states, these transitions have non-exponential waiting times.

\subsection{Master equation}
Kinetic models with non-exponential waiting times are known as semi-Markov processes \cite{Gillespie77,Esposito08,Maes09}.
Let us denote by $P(i,t)$ the (conditional) probability that the motor is in state $i$ at time $t$ given that it arrived in state $1$ at $t=0$. In a semi-Markov process we need to introduce for each transition a waiting time density $\psi(t)$. In our model, we will assume that only the transition between the states $M_2$ and $M_3$ is non-Markovian. As explained in the previous subsection this should be a proper description of the motion of the tethered head through a viscoelastic medium like the cellular cytoplasma. We will denote the associated waiting time density by $\psi_S(t)$. The other three transition, $M_1 \to M_2$, $M_2 \to M_1$ and $M_3 \to M_1$ will be assumed to be Markovian with associated transition rates $k_+, k_-$ and $k_3$ respectively. The associated waiting time distributions, $\psi_{21}, \psi_-$ and $\psi_{13}$ are therefore exponential functions.

An extra complication arises because from the state 2 we can go either to state 1 or state 3. In order to make the transition from 2 to 1 in the interval $[t,t+dt]$ we should not have made the transition between 2 and 3 at an earlier time. Therefore the waiting time distribution $\psi_{12}(t)$ to go from 2 to 1 {\it and not to 3} equals
\begin{eqnarray}
\psi_{12}(t) = \psi_-(t)\ \int_t^\infty \psi_S(t')\ dt'
\label{1}
\end{eqnarray}
Analogously, the waiting time distribution $\psi_{32}(t)$ to go from 2 to 3 and not to 1 is given by
\begin{eqnarray}
\psi_{32}(t) = \psi_S(t)\ \int_t^\infty \psi_-(t')\ dt'
\label{2}
\end{eqnarray}

Using the theory of semi-Markov processes \cite{Gillespie77,Esposito08} we can now write down the master equation for our model which is a set of linear integro-differential equations
\begin{widetext}
\begin{eqnarray}
\frac{\partial P(1,t)}{\partial t} &=&  \int_0^t \left[w_{12}(t') P(2,t-t') + w_{13} (t') P(3,t-t') - w_{21}(t') P(1,t-t')\right]dt' \nonumber\\
\frac{\partial P(2,t)}{\partial t} &=&  \int_0^t \left[w_{21}(t') P(1,t-t') - (w_{32} (t') + w_{12}(t')) P(2,t-t')\right] dt'\nonumber \\
\frac{\partial P(3,t)}{\partial t} &=&  \int_0^t \left[w_{32}(t') P(2,t-t') - w_{13}(t') P(3,t-t')\right]dt'
\label{3}
\end{eqnarray}
\end{widetext}
The functions $w_{ij}(t)$ can be expressed in terms of the waiting time distributions $\psi_{ij}(t)$. The precise relation will be given in the next section. The initial condition is $P(i,0)=\delta_{i1}$. 

The quantity of interest is the position $x(t)$ of the motor. Each time the motor makes the transition between the states 2 and 3 its position increases with $d_0 \approx 4.1$ nm. The position is a stochastic variable and in section 4 we will explain how to obtain all  its moments $\langle x(t)^n\rangle$. From the first moment we can get the average velocity 
\begin{eqnarray}
V = \lim_{t \to \infty} \frac{d}{dt} \langle x(t) \rangle
\label{vel}
\end{eqnarray}
Indeed, we will show that in our model the motor will always move ballistically. Similarly, the dispersion around the average position always has a diffusive behaviour. The associated diffusion constant is given by
\begin{eqnarray}
D = \frac{1}{2} \lim_{t \to \infty} \frac{d}{dt} \left[ \langle x^2(t) \rangle - \langle x(t)\rangle^2 \right]
\label{diff}
\end{eqnarray}
In principle higher moments can be calculated and one can also determine, for example, the skewness or a parameter that quantifies deviations from Gaussianity. However as we will see below the calculations become quite involved and we limit ourselves in this paper to the first two moments.

\section{The average velocity of the motor}
\subsection{General results}
Our first aim is to calculate the average velocity $V$ of the motor. 

Since the integrals on the right hand side of the master equation (\ref{3}) are convolutions, it is convenient to do a Laplace transform. Let us denote by $\tilde{f}(s)=\int_0^\infty f(t) e^{-st} dt$ the Laplace transform of a function $f(t)$. Then (\ref{3}) can be rewritten as
\begin{widetext}
\begin{eqnarray}
s\tilde{P}(1,s) &=& \tilde{w}_{12}(s) \tilde{P}(2,s) + \tilde{w}_{13}(s) \tilde{P}(3,s) - \tilde{w}_{21}(s) \tilde{P}(1,s) + 1 \nonumber \\
s\tilde{P}(2,s) &=& \tilde{w}_{21}(s) \tilde{P}(1,s) - (\tilde{w}_{32}(s) + \tilde{w}_{12}(s)) \tilde{P}(2,s) \nonumber \\
s\tilde{P}(3,s) &=& \tilde{w}_{32}(s) \tilde{P}(2,s) - \tilde{w}_{13}(s) \tilde{P}(3,s)
\label{4}
\end{eqnarray}
\end{widetext}

According to the general formalism of semi-Markov processes (for a clear explanation, see \cite{Esposito08}), the Laplace transform $\tilde{w}_{ij}(s)$ can be related to those of the waiting time distributions $\tilde{\psi}_{ij}(s)$. The explicit relations depend on the reaction scheme and for the present case are given by
\begin{eqnarray}
\tilde{w}_{12}(s) &=& \frac{s \tilde{\psi}_{12}(s)}{1- \tilde{\psi}_{12}(s) - \tilde{\psi}_{32}(s)} \nonumber \\
\tilde{w}_{21}(s) &=& \frac{s \tilde{\psi}_{21}(s)}{1-\tilde{\psi}_{21}(s)} \nonumber \\
\tilde{w}_{13}(s) &=& \frac{s \tilde{\psi}_{13}(s)}{1-\tilde{\psi}_{13}(s)} \nonumber \\
\tilde{w}_{32}(s) &=& \frac{s \tilde{\psi}_{32}(s)}{1-\tilde{\psi}_{12}(s) - \tilde{\psi}_{32}(s)} \label{5}
\end{eqnarray}

Assuming as we did that the waiting time distributions $\psi_{21}(t), \psi_-(t)$ and $\psi_{13}(t)$ are exponential with rates $k_+, k_-$ and $k_3$ and using the relations (\ref{ws}), (\ref{1}) and (\ref{2}) one obtains immediately
\begin{eqnarray}
\tilde{w}_{12}(s) &=& k_- \nonumber \\
\tilde{w}_{21}(s) &=& k_+ \nonumber \\
\tilde{w}_{13}(s) &=& k_3 \nonumber \\
\tilde{w}_{32}(s) &=& \left[ \tilde{S}(s+k_-)\right]^{-1} -s - k_-
\label{6}
\end{eqnarray}
where $\tilde{S}(s)$ is the Laplace transform of the survival probability $S(t)$. For the moment we want to obtain results for arbitrary survival times and therefore do not specify an explicit form for $S(t)$.

After substitution of the expressions (\ref{6}), solution of the set of linear equations (\ref{4}) gives $\tilde{P}(i,s)$.  

In order to obtain the average velocity we observe that the motor makes a step each time the transition from 2 to 3 is made. Therefore, the average number of steps made by the motor per unit of time is given by the current
\begin{eqnarray}
J(t) = \int_0^t w_{32}(t') P(2,t-t')\ dt'
\label{11}
\end{eqnarray}
The average velocity at time $t$, $V(t)$ clearly is given by $d_0 J(t)$. 

Again it is convenient to look first at the Laplace transforms of equation (\ref{11}) which reads\begin{eqnarray}
\tilde{J}(s) = \tilde{w}_{32}(s) \tilde{P}(2,s)
\label{13}
\end{eqnarray}

Inserting the solution for $\tilde{P}(2,s)$ one obtains for the Laplace transform of the current an expression in terms of the transition rates $k_+, k_-, k_3$ and the survival probability $\tilde{S}(s)$. We find
\begin{eqnarray}
\tilde{J}(s) = \frac{\left(1-(k_-+s)\tilde{S}(s+k_-)\right)\left(k_+(k_3+s)\right)}{s\left(s+k_3+k_++(k_3-k_-)k_+ \tilde{S}(s+k_-)\right)}
\label{14}
\end{eqnarray}

It follows that, independently of the precise form of the survival probability $S(t)$, the motor will always move ballistically in the long time limit.
Indeed, in the $s \to 0$ limit of (\ref{14}) one gets
\begin{eqnarray}
\lim_{s\to 0}  s\tilde{J}(s) =  \left[\frac{\left(1-k_- \tilde{S}(k_-)\right) k_+ k_3}{k_3 + k_+ +(k_3-k_-)k_+ \tilde{S}(k_-)}\right]
\label{14pr}
\end{eqnarray}
The final value theorem for Laplace transforms then implies that for $t\to \infty$ the current will be a constant that is equal to the right hand side of (\ref{14pr}). Hence, the asymptotic velocity of the motor (\ref{vel}) equals
\begin{eqnarray}
V=d_0 \left[\frac{\left(1-k_- \tilde{S}(k_-)\right) k_+ k_3}{k_3 + k_+ +(k_3-k_-)k_+ \tilde{S}(k_-)}\right]
\label{20}
\end{eqnarray}

The distance travelled by the motor increases linearly in time, and is not changed to, for example, subballistic motion.  

The step between the states 1 and 2 depends on the concentration, $[\mbox{ATP}]$, of ATP. Hence we write $k_+=k_0\ [\mbox{ATP}]$. Inserting this in (\ref{20}) shows that the velocity of the motor is always of the Michaelis-Menten form
\begin{eqnarray}
V = \frac{V_m\ [\mbox{ATP}]}{K_M + [\mbox{ATP}]}
\label{21}
\end{eqnarray}
where the maximum velocity of the motor $V_m$ is given by
\begin{eqnarray}
V_m= d_0 \left[\frac{\left(1-k_- \tilde{S}(k_-)\right)  k_3}{1 +(k_3-k_-) \tilde{S}(k_-)}\right]
\label{22}
\end{eqnarray}
while the Michaelis-Menten constant $K_M$ equals
\begin{eqnarray}
K_M = \frac{k_3}{\left[1 + (k_3-k_-) \tilde{S}(k_-)\right] k_0}
\label{23}
\end{eqnarray}
This is our first main result. It is consisted with the experiments of \cite{Sozanski15} where it was found that the motion of kinesin in the presence of crowders is ballistic and that the velocity has the Michaelis-Menten form. In contrast, in the ratchet model of \cite{Goychuk14b} both ballistic and subballistic motion was found. However in that work the motor was coupled to a cargo that gives rise to an extra force on the motor. A cargo is absent in our calculations and in the experiments of \cite{Sozanski15}.


\subsection{Stretched exponential survival time} 
As argued in section 2, a good choice for the survival probability is a stretched exponential (\ref{se}) in which case the waiting time distribution becomes Weibull.
No closed expressions for the Laplace transform $\tilde{S}(s)$ of a stretched exponential is known. However, the transform can easily be performed numerically.

For the parameters $k_0, k_-$ and $k_3$ we have taken the experimentally determined values in \cite{Clancy11} : $k_0 = 3.7 \mu$M$^{-1}$ s$^{-1}$, $k_-=68$ s$^{-1}$ and $k_3=57$ s$^{-1}$. The stretched exponential function is characterised by two parameters $k_s$ and $\beta$. For $\beta=1$, we recover an exponential waiting time with rate $k_s$. Our model then corresponds to the standard Markov model for which $k_S=570$ s$^{-1}$ experimentally \cite{Clancy11}. 

\begin{figure}[h]
\centering
\includegraphics[width=8cm]{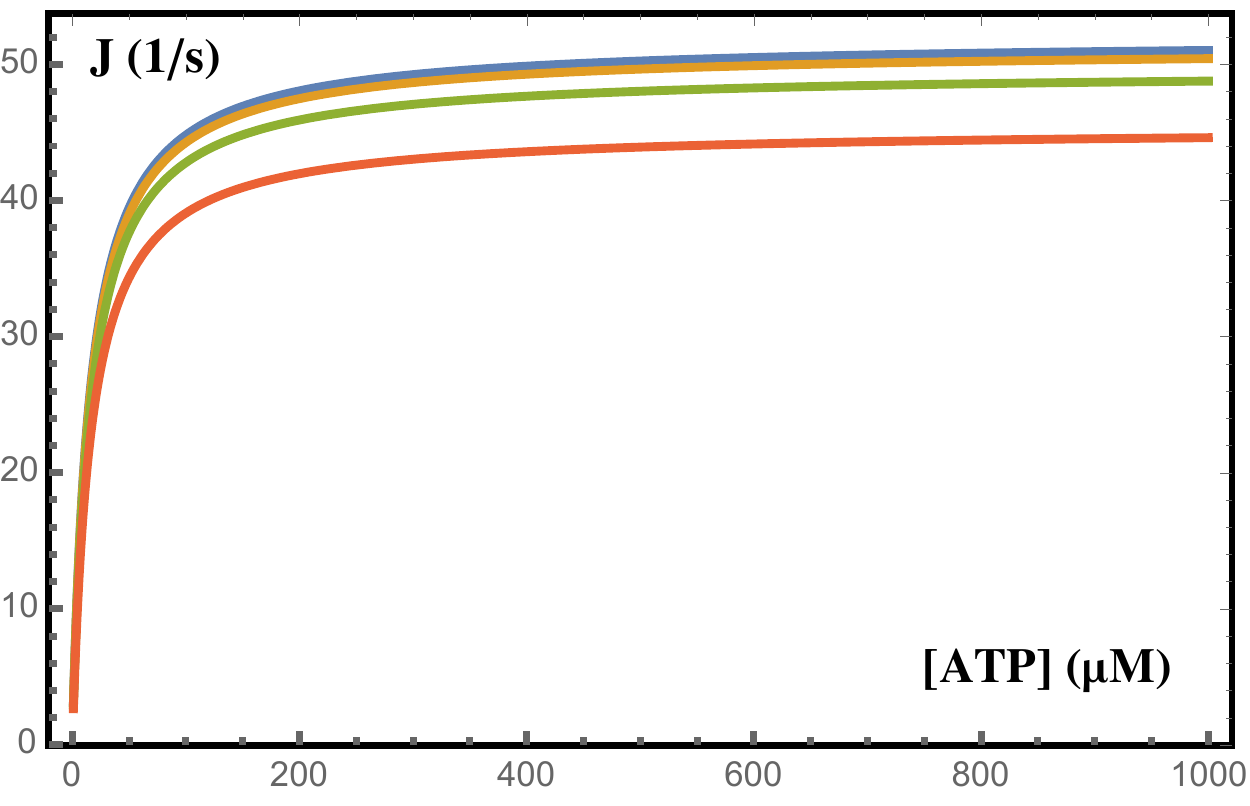}
\caption{Michaelis-Menten plot of the motor current $J=V/d_0$ for $\beta=1, .75, .5$ and $0.25$ (top to bottom) and $k_S=570$ s$^{-1}$.}
\label{Fig3}
\end{figure}

In Fig.\ \ref{Fig3} we plot the velocity of the motor as a function of ATP-concentration for $k_S=570$ s$^{-1}$ and four values of $\beta$.

We see that the maximum velocity of the motor decreases with $\beta$. The precise relation between the exponents $\beta$ of the stretched exponential and $\alpha$ of the viscoelastic environment is not known but the simulations of \cite{Goychuk09} show that $\beta$ is an increasing function of $\alpha$. The maximum velocity of the motor therefore decreases as the medium becomes less viscous and more elastic. 
This picture is at least qualitatively in agreement with the experiments of \cite{Sozanski15} where it was found that in the presence of the crowder sucrose, the maximum velocity of kinesin-1 decreased with a factor two. 

\begin{figure}[h]
\centering
\includegraphics[width=8cm]{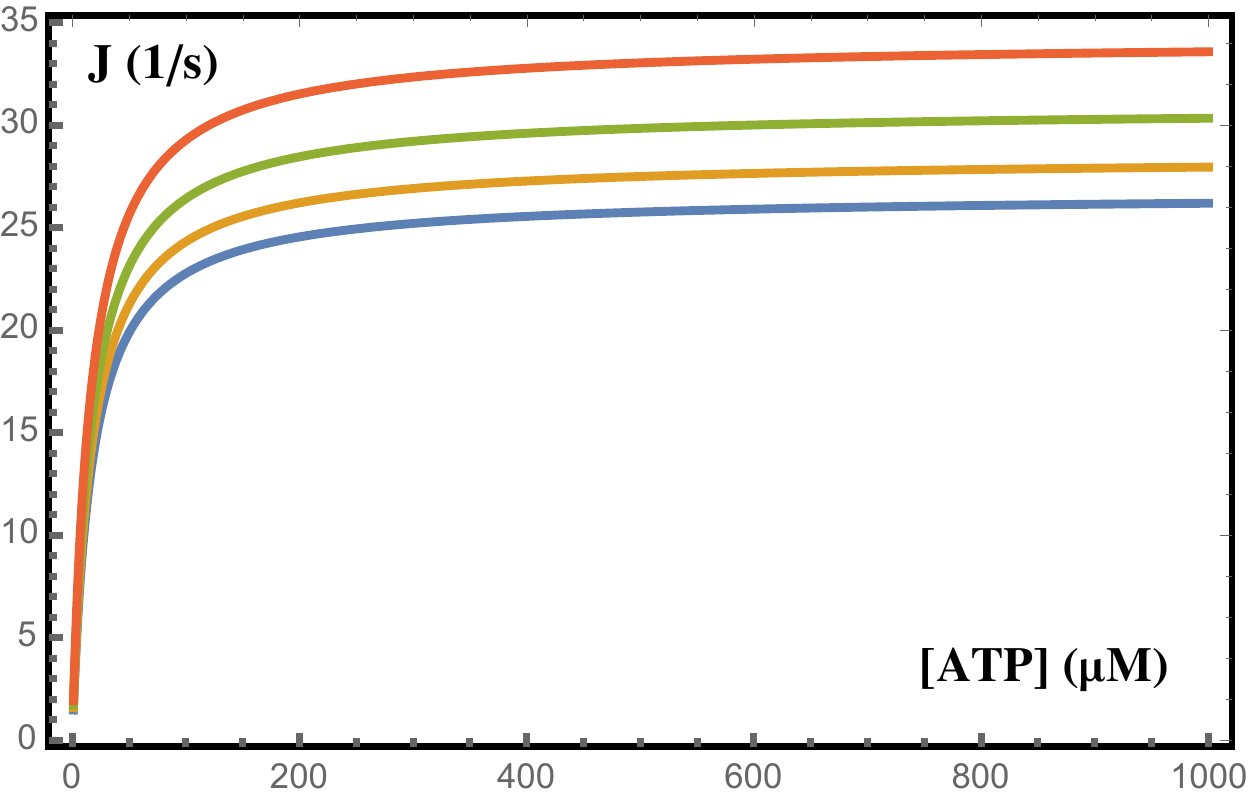}
\caption{Michaelis-Menten plot of the motor current $J$ (steps per second) for $\beta=1, .75, .5$ and $0.25$ (bottom to top) and $k_S=50$ s$^{-1}$.}
\label{Fig4}
\end{figure}

For values of $k_s$ close to $100$ s$^{-1}$ the velocity has a very weak dependence on $\beta$  while for
small values of $k_s$ ($k_s < 100$ s$^{-1}$) the trend is reversed: the less elastic the medium, the slower the motor moves (see Fig.\ \ref{Fig4}). It is however unclear whether this regime is ever encountered in biological situations.

\section{Moment generating function of the position}
Next, we want to determine the second moment of the motor's position, i.e. $\langle x^2(t) \rangle$ which allows us to determine the diffusion constant $D$ using (\ref{diff}). The standard technique to calculate $D$ for molecular motors \cite{Fisher99,Fisher99b} cannot be easily extended to non-exponential waiting times. Here we therefore follow an alternative route in which we first determine the full moment generating function  of the position. 

It is convenient to measure the position of the motor in terms of the number of steps $n_x(t)$ made, with the obvious relation $x(t) = d_0 n_x(t)$. We are interested in the moment generating function of $n_x(t)$, i.e. in $G(q,t)=\langle e^{q n_x(t)} \rangle$.
The $m$-th moment of $x(t)$ can be found in terms of the $m$-th order coefficient in a Taylor expansion of $G(q,t)$
\begin{eqnarray}
\langle n_x^m(t) \rangle = \frac{d^m G(q,t)}{d q^m}(q=0)
\label{23p}
\end{eqnarray}

 The moment generating function and the related cumulant generating function play an important role in the large deviation approach to (non-equilibrium) statistical mechanics \cite{Touchette09}. Within that context, it has been shown that $G(q,t)$ obeys a modified master equation. This was first shown for Markov processes in \cite{Lebowitz99}, and was then extended to semi-Markov processes in \cite{Esposito08,Maes09}.  

For completeness, we briefly outline the derivation of this modified master equation. Let $P(i,n_x,t)$ be the probability that the motor is in the state $i$ and has made $n_x$ steps at the moment $t$. In the present model, $n_x$ can only be modified when the transition form state $2$ to state $3$ is made. Hence, one immediately realises that $P(i,n_x,t)$ obeys
\begin{widetext}
\begin{eqnarray*}
\frac{\partial P(1,n_x,t)}{\partial t} &=&  \int_0^t \left[w_{12}(t') P(2,n_x,t-t') + w_{13} (t') P(3,n_x,t-t') - w_{21}(t') P(1,n_x,t-t')\right]dt' \nonumber\\
\frac{\partial P(2,n_x,t)}{\partial t} &=&  \int_0^t \left[w_{21}(t') P(1,n_x,t-t') - (w_{32} (t') + w_{12}(t')) P(2,n_x,t-t')\right] dt'\nonumber \\
\frac{\partial P(3,n_x,t)}{\partial t} &=&  \int_0^t \left[w_{32}(t') P(2,n_x-1,t-t') - w_{13}(t') P(3,n_x,t-t')\right]dt'
\end{eqnarray*}
\end{widetext}

Next, we introduce the discrete Laplace transform of $P(i,n_x,t)$ with respect to $n_x$
\begin{eqnarray}
P_L(i,q,t) \equiv \sum_{n_x} e^{q n_x} P(i,n_x,t)
\label{24}
\end{eqnarray}
A simple calculation shows that $P_L(i,q,t)$ obeys almost the same master equation as $P(i,t)$. The only difference occurs in the equation for $P_L(3,q,t)$ where in the first term on the right hand side one needs to replace $w_{32}(t')$ by $w_{32}(t') e^q$ 
\begin{widetext}
\begin{eqnarray}
\frac{\partial P_L(1,q,t)}{\partial t} &=&  \int_0^t \left[w_{12}(t') P_L(2,q,t-t') + w_{13} (t') P_L(3,q,t-t')- w_{21}(t') P_L(1,q,t-t')\right]dt' \nonumber\\
\frac{\partial P_L(2,q,t)}{\partial t} &=&  \int_0^t \left[w_{21}(t') P_L(1,q,t-t') - (w_{32} (t') + w_{12}(t')) P_L(2,q,t-t')\right] dt'\nonumber \\
\frac{\partial P_L(3,q,t)}{\partial t} &=&  \int_0^t \left[w_{32}(t') e^q P_L(2,q,t-t') - w_{13}(t') P_L(3,q,t-t')\right]dt'
\label{25}
\end{eqnarray}
\end{widetext}
Finally, we observe that the generating function obeys
\begin{eqnarray}
G(q,t) = \sum_{i=1}^3 \sum_{n_x} e^{q  n_x} P(i,n_x,t)= \sum_{i=1}^3 P_L(i,q,t)
\label{26}
\end{eqnarray} 
Hence, after solving (\ref{25}), we can obtain the various moments of $n_x$ using (\ref{26}) and (\ref{23p}). 

In order to solve (\ref{25}), we go over to the Laplace transforms with respect to time, $\tilde{P}_L(i,q,s)$, of $P_L(i,q,t)$. If we moreover assume that at $t=0$ the motor is in the state $i=1, n_x=0$ we obtain the following linear set of equations for $\tilde{P}_L(i,q,s)$
\begin{widetext}
\begin{eqnarray}
s\tilde{P}_L(1,q,s) &=& \tilde{w}_{12}(s) \tilde{P}_L(2,q,s) + \tilde{w}_{13}(s) \tilde{P}_L(3,q,s) - \tilde{w}_{21}(s) \tilde{P}_L(1,q,s) + 1 \nonumber \\
s\tilde{P}_L(2,q,s) &=& \tilde{w}_{21}(s) \tilde{P}_L(1,q,s) - (\tilde{w}_{32}(s) + \tilde{w}_{12}(s)) \tilde{P}_L(2,q,s) \nonumber \\
s\tilde{P}_L(3,q,s) &=& \tilde{w}_{32}(s) e^q \tilde{P}_L(2,q,s) - \tilde{w}_{13}(s) \tilde{P}_L(3,q,s)
\label{27}
\end{eqnarray}
\end{widetext}
In Laplace space, the relation (\ref{26}) becomes
\begin{eqnarray}
\langle \int_0^\infty e^{-st} e^{q n_x(t)} dt\rangle= \sum_{i=1}^3 \tilde{P}_L(i,q,s)
\end{eqnarray}
so that after doing a Taylor expansion of this relation in the variable $q$ we obtain
\begin{eqnarray}
\langle \tilde{n}_x^m(s) \rangle = \frac{d^m (\sum_{i=1}^3 \tilde{P}_L(i,q,s))}{dq^m}(q=0)
\end{eqnarray}
Finally, inverse Laplace transforms then allow us to find the moments of the position of the motor.

The necessary calculations can easily be done using Mathematica and give the following results
\begin{eqnarray}
\hspace*{-1cm}\langle \tilde{n}_x(s) \rangle= 
\frac{\left(1-(k_-+s)\tilde{S}(s+k_-)\right)\left(k_+(k_3+s)\right)}{s^2\left(s+k_3+k_++(k_3-k_-]k_+ \tilde{S}(s+k_-)\right)}
\label{28} \\
\hspace*{-1cm}\langle \tilde{n}_x^2(s) \rangle=\frac{k_+(k_3+s)\left(-1+(k_-+s) \tilde{S}(s+k_-)\right)T(s)}{s^3\left(k_3 + k_+ + s+k_+(k_3-k_-) \tilde{S}(s+k_-)\right)^2}
\label{29}
\end{eqnarray}
where
\begin{eqnarray*}
T(s)&=&-k_3 s -s \left(k_+ +s - k_- k_+ \tilde{S}(s+k_-)\right) \\ &+& k_3 k_+  \left(-2 + (2 k_- +s) \tilde{S}(s+k_-)\right)
\end{eqnarray*}

Notice that the current $J(t)$ introduced earlier is nothing but the time derivative of $\langle n_x(t) \rangle$. In Laplace space this implies $\langle \tilde{n}_x(s) \rangle = \tilde{J}(s)/s$, a relation that is indeed satisfied (compare (\ref{14}) with (\ref{28})). 

\section{The diffusion constant}
\subsection{General results}
The diffusion constant can be determined from the first and the second moment of $n_x(t)$.
Unfortunately it is not possible to calculate these from their Laplace transforms (\ref{28}) and (\ref{29}) for arbitrary times. However, in order to determine the diffusion constant we need only the behaviour of these functions for $t \to \infty$ which can again be obtained using the final value theorem for Laplace transforms. 

We therefore first make Laurent expansions of (\ref{28}) and (\ref{29}) around $s=0$. The results are of the form
\begin{eqnarray*}
\langle \tilde{n}_x(s) \rangle = \frac{A}{s^2} + \frac{B}{s} + \cdots \\
\langle \tilde{n}_x^2 (s) \rangle = \frac{2A^2}{s^3} + \frac{C}{s^2} + \cdots 
\end{eqnarray*}
where the coefficients $A, B, C, ...$ are complicated functions of the transition rates $k_, k_-, k_3$. They also depend on the (Laplace transform of the) waiting time distribution $\tilde{S}(s)$. For example, $A$ is given by the right hand side of (\ref{14pr}). 
From the above relations  we find that
\begin{eqnarray*}
\lim_{s \to 0} s (\langle \tilde{n}_x (s) \rangle-A/s^2) = B \\
\lim_{s \to 0} s (s\langle \tilde{n}_x^2 (s)\rangle -2A/s^2 )=C
\end{eqnarray*}
The inverse Laplace transforms of the terms between brackets on the left hand side of these equations can immediately be found. Using
the final value theorem it follows that
\begin{eqnarray}
\lim_{t \to \infty} (\langle n_x(t) \rangle - At) = B \label{30}\\
\lim_{t \to \infty} (d\langle n_x^2(t) \rangle/dt - 2A^2 t)=C
\label{31}
\end{eqnarray}
since $d\langle n_x(t)\rangle/dt (0)$ is finite. Combining these results, we obtain
\begin{eqnarray*}
\lim_{t \to \infty} \frac{\langle n_x^2 (t) \rangle - \langle n_x(t) \rangle^2}{t}= (C-2B A)
\end{eqnarray*}
so that we get the final expression for the diffusion coefficient
\begin{eqnarray}
D = d_0^2 (C-2BA)/2
\label{32}
\end{eqnarray}

The coefficients $A, B$ and $C$ in the Laurent expansions  of (\ref{28}) and (\ref{29}) were calculated using Mathematica. This leads to the following general expression for the diffusion coefficient of our model:
\begin{eqnarray}
D&=&d_0^2 \Big((k_3 k_+ (1-k_- \tilde{S}(k_-))(k_3^2+k_+^2 -2 k_-k_+^2 \tilde{S}(k_-) \nonumber \\ &+& (-k_3^2+k_-^2)k_+^2 \tilde{S}(k_-)^2 - 2 k_3^2 k_+ (k_-+k_+) d\tilde{S}/ds(k_-))\Big)/ \nonumber \\
& & \Big( 2 (k_3 + k_+ + (k_3-k_-) k_+ \tilde{S}(k_-))^3\Big)
\label{33}
\end{eqnarray}

In conclusion, we find that asymptotically in time, the motor always diffuses with the diffusion constant given by (\ref{33}). Hence, there is no subdiffusion or superdiffusion of the motor. This is in contrast to passive diffusion, where viscoelasticity changes diffusive behaviour into subdiffusive motion. Our results do not rule out that there can be an early time regime in which such behaviour is present. However, simulations of our model with various waiting times showed that the asymptotic regime is almost reached immediately and also gave no evidence for super- or subdiffusion. We therefore concentrate on the behaviour of the diffusion constant (\ref{33}) as a function of the various parameters in the model.

\subsection{Stretched exponential survival time}
We have numerically evaluated the expression (\ref{33}) for the parameter values quoted in subsection 3.2 both as a function of ATP-concentration and as a function of the parameters in the Weibull distribution.

\begin{figure}[h]
\centering
\includegraphics[width=8cm]{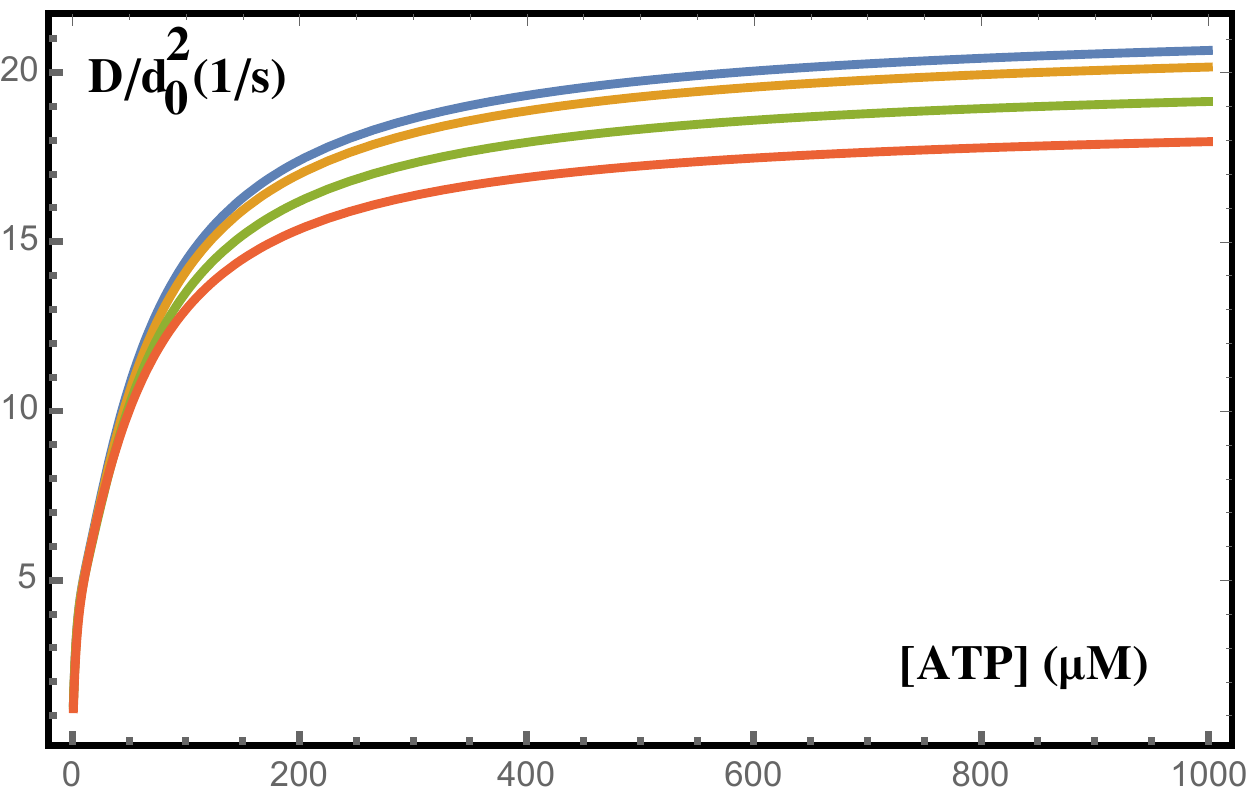}
\caption{Diffusion constant $D/d_0^2$ of the motor as a function of ATP-concentration  for $\beta=1, .75, .5$ and $0.25$ (top to bottom) and $k_S=570$ s$^{-1}$.}
\label{Fig5}
\end{figure}

\begin{figure}[h]
\centering
\includegraphics[width=8cm]{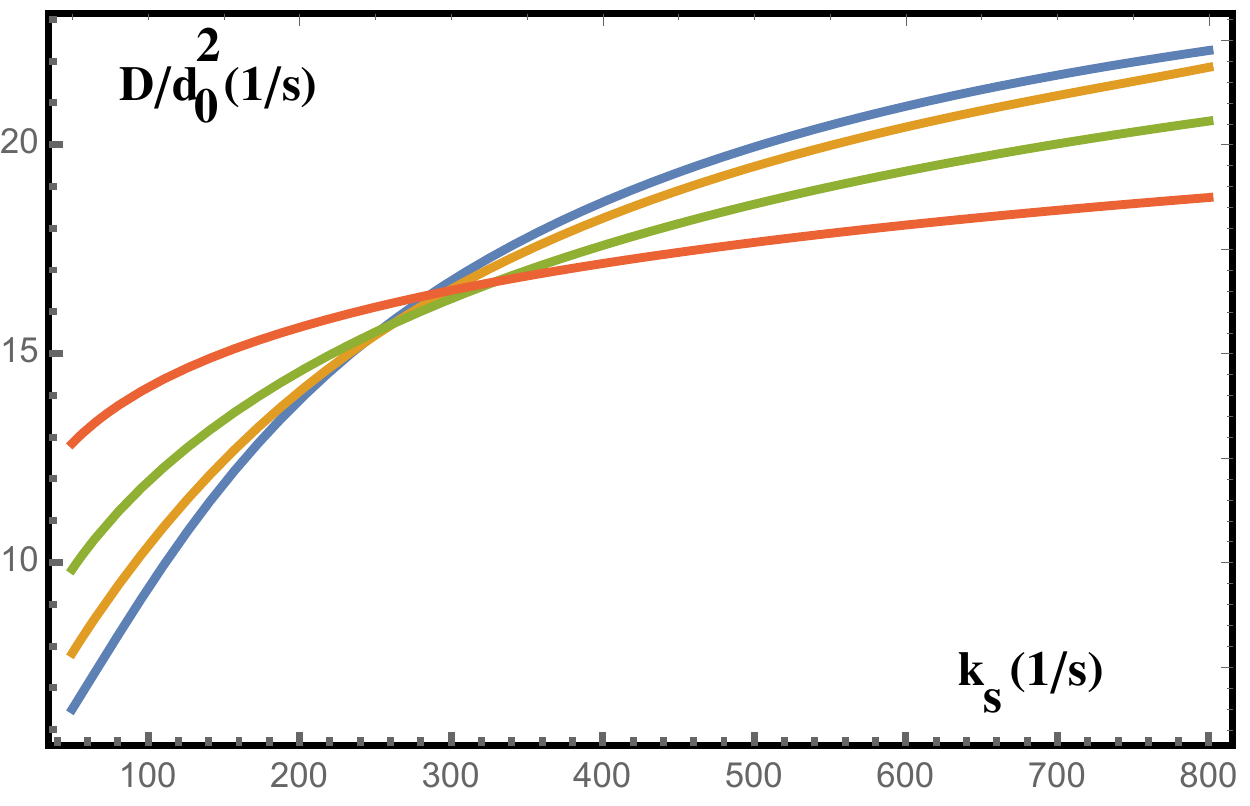}
\caption{Diffusion constant $D/d_0^2$ of the motor as a function of $k_S$  for $\beta=1, .75, .5$ and $0.25$ (blue, orange, green and red curve respectively) and $[\mbox{ATP}]=1000 \mu$M.}
\label{Fig6}
\end{figure}

In Fig.\ \ref{Fig5} we plot the diffusion constant of the motor as a function of ATP-concentration for $k_S=570$ s$^{-1}$ and four values of $\beta$. 
As was the case for the velocity, the diffusion constant decreases if $\beta$ decreases, i.e. when the environment becomes less viscous and more elastic. Again, for small enough values of $k_S$, this trend reverses.  In Fig.\ \ref{Fig6} we plot the diffusion constant as a function of $k_S$ for four $\beta$-values and an ATP-concentration of $1000\ \mu$M. We see that while for each value of $\beta$, $D$ decreases with decreasing $k_S$ this effect is much stronger for higher $\beta$. Assuming that $k_S$ is still inversely proportional to the viscosity, this result implies that as the medium becomes more elastic, the dependence of the motor's diffusion constant on viscosity becomes weaker. 

\section{Fano factor}
From the asymptotic results on the diffusion constant and the velocity we can obtain the Fano factor $F$, also sometimes called randomness parameter \cite{Kolomeisky07}, which is given by
\begin{eqnarray}
F \equiv \frac{\langle n_x^2 \rangle - \langle{n_x}\rangle^2}{\langle n_x \rangle}=\frac{2 D}{d_0^2 J}
\label{34}
\end{eqnarray}

For enzymatic reactions it is a quantity that can be measured in single molecule experiments and that can give information on the number of kinetic states $N$ occuring in the reaction. Indeed, it has been shown that
$F \geq 1/N$ \cite{Moffitt10,Moffitt14}. More recently sharper bounds on this quantity have been obtained by considering thermodynamic constraints on the cost of generating fluctuations in, for example,  the number of steps made by a motor or the number of molecules consumed by an enzyme \cite{Barato15,Pietzonka17,Proesmans17}. These results were obtained for Markov processes. We are not aware of similar bounds for semi-Markov processes. It would be interesting if measurements of the randomness could give information on the importance of non-Markovian effects in reaction kinetics.

\begin{figure}[h]
\centering
\includegraphics[width=8cm]{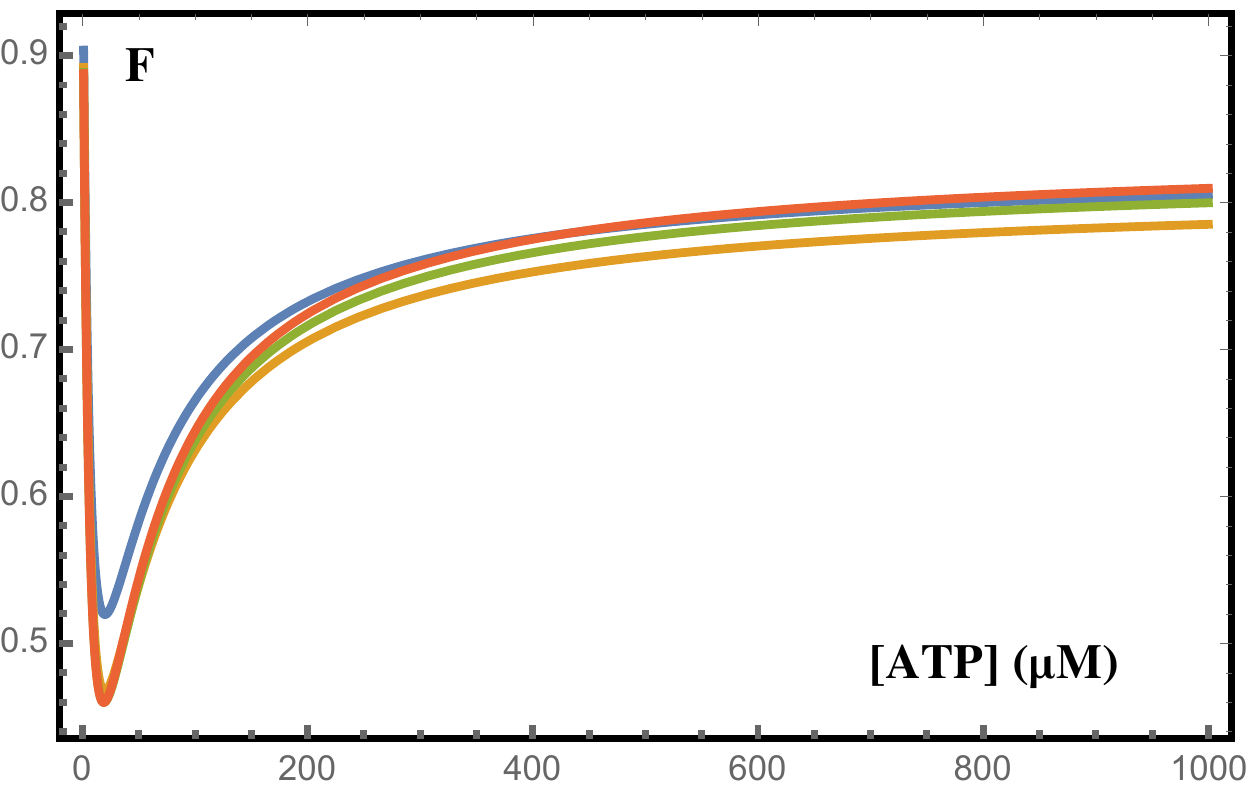}
\caption{Fano factor $2D/d_0^2 J$ of the motor as a function of ATP-concentration  for $\beta=1, .75, .5$ and $0.25$ (bottom to top) and $k_S=570$ s$^{-1}$.}
\label{Fig7}
\end{figure}

\begin{figure}[h]
\centering
\includegraphics[width=8cm]{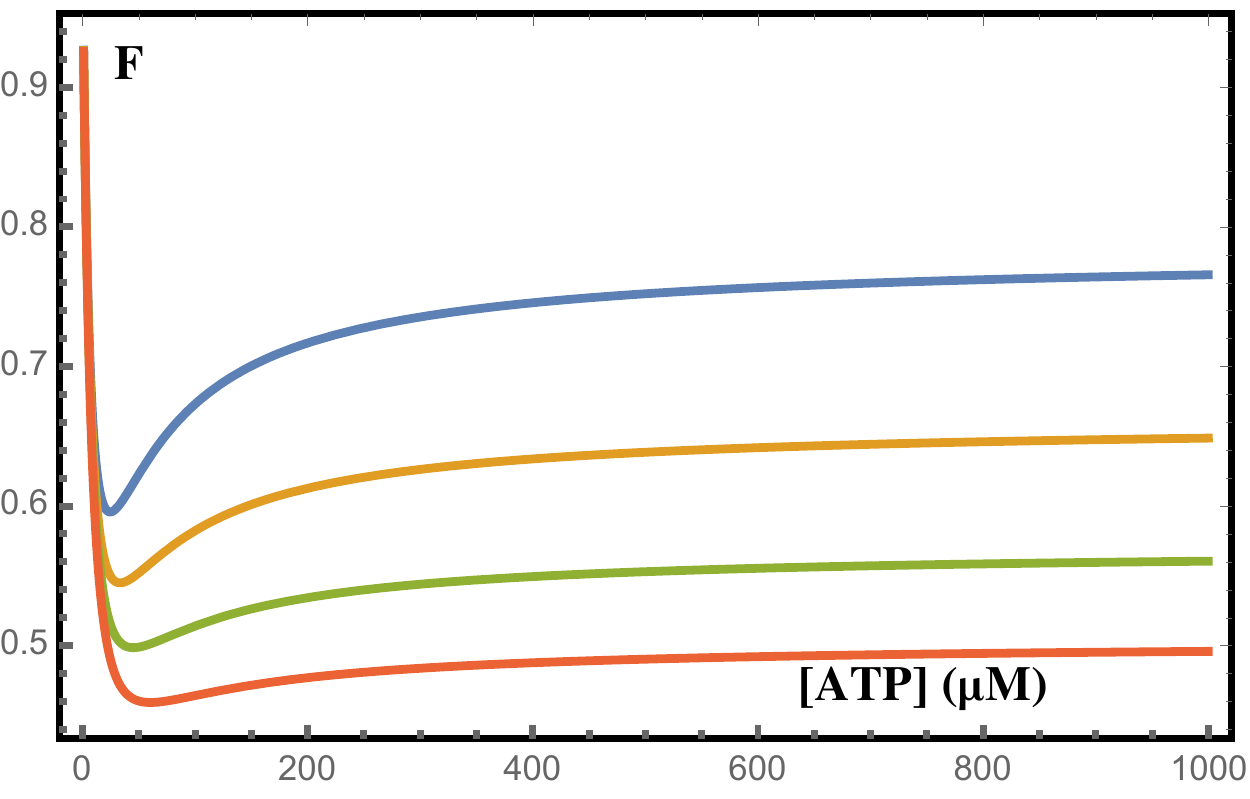}
\caption{Fano factor $2D/d_0^2 J$ of the motor as a function of ATP-concentration  for $\beta=1, .75, .5$ and $0.25$ (bottom to top) and $k_S=50$ s$^{-1}$.}
\label{Fig8}
\end{figure}

We calculated the Fano factor for our semi-Markov model with a stretched exponential survival time and for the parameter values quoted in section 3.2. Since $N=3$ in our model, the Markovian bound is $F \geq 1/3$. In Fig.\ \ref{Fig7} we plot the Fano factor  as a function of ATP-concentration for $k_S=570$ s$^{-1}$ and four values of $\beta$. 
Similar results are shown for $k_S=50$ s$^{-1}$ in Fig.\ \ref{Fig8}.
The bound $F \geq 1/3$ is alvvays satisified. In these figures we also see that the Markovian result  lies below the semi-Markov one. Our results show that care needs to be taken in interpreting Fano factors in the presence of non-Markovian effects. Indeed, for the parameter values of Fig.\ \ref{Fig7} and for $\beta=0.25$ one observes that $F \geq 1/2$ which could lead to the conclusion that $N=2$ if one assumes that the underlying process is Markov. 
It would therefore certainly be interesting to see whether recent bounds on the Fano factor \cite{Barato15} can be extended to the semi-Markov case.

\section{Conclusions}
In this paper we have investigated how the viscoelasticity of the cytosol influences the motion of the molecular motor kinesin. We have argued that in such a medium the tethered motorhead experiences memory-dependent friction. In the description of the motor motion in terms of a kinetic model this leads to a semi-Markov model where the step in which the motor moves forward has a non-exponential waiting time. 

We have determined expressions for both the velocity (\ref{20}) and the diffusion constant (\ref{33}) of the motor for arbitrary waiting time distributions. We introduced a calculational technique  that allows us to calculate the full generating function of the position of the motor.  We have found that the motion of the motor always remains regular, i.e. that is moves with a constant velocity and that the dependence on ATP-concentration is of Michaelis-Menten form. The spreading of the position around its average value is diffusive. 

These results are general and do not depend on the waiting time distribution.
It would be interesting if in an experiment one could measure the time between the binding of ATP to the tethered head and the subsequent binding of that head to the microtubule. Measuring the distribution of that time would provide the necessary input for our model. Since such measurements are not available at the moment we argued on the basis of existing theoretical work that a Weibull distribution can be a reasonable approximation. The Weibull distribution depends on two parameters: $\beta$ which can be related to the rheological parameter $\alpha$ and $k_s$ which, at least in the viscous case, is inversely proportional to the viscosity of the medium. We have calculated the velocity and the diffusion constant of our model for a Weibull waiting time density. For realistic parameter values both quantities decrease if the medium becomes more elastic and less viscous (as measured by the parameters $\alpha$ or $\beta$). However, the dependence of $V$ and $D$ on $\beta$ is rather weak if $\beta$ is not too different from 1. This could possibly explain why motor properties are not too much influenced by viscoelasticity \cite{Ross16}. 

We have also calculated the Fano factor or randomness, a quantity which received a lot of recent interest in the context of the thermodynamic uncertainty relations \cite{Barato15,Pietzonka17,Proesmans17}. We observe that also in the present model,  $F$ is bounded from below by $1/N$, where $N=3$ is the number of kinetic states in the model. However, it would certainly be of interest to investigate the validity of the thermodynamic uncertainty relations more generally for semi-Markov processes and to see whether statistical kinetics can be used to get insight on the role of memory effects in the motion of motors or in other enzymatic reactions. 

In the present work we did not include the effect of a force acting on the motor. This force can be due to a cargo or can be exerted artificially with an optical tweezer. The presence of a force will modify the waiting time distribution $\psi_S(t)$ as it does in the viscous case. At this moment we have no clear idea what will be the precise effect: will it modify the functional form of $\psi_S(t)$, or will it remain Weibull and will only the parameters $\beta$ and $k_s$ be changed? Indeed for the viscous case, it is known that only the rate $k_S$ is dependent on the applied force. However, on the basis of our general results we believe that in the presence of a force, the motor will still move with a constant velocity that will approach zero at some stalling force. Within the present scenario it is not possible to obtain subballistic or subdiffusive motion.

{\bf Acknowledgement} We thank Stefanie Put for help with the figures.


\begin{thebibliography}{99}
\bibitem{Clancy11}
Clancy B E, Behnke-Parks W M, Andreasson J O L, Rosenfeld S S and Block S M  {\it Nat. Struct. Mol. Biol.} {\bf 18} 1020 (2011).
\bibitem{Howard01}
Howard J  {\it Mechanics of motor proteins and the cytoskeleton}  Sinauer (2001).
\bibitem{Svoboda93} 
Svoboda K, Schmidt C F, Schnapp B J and Block S M  {\it Nature} {\bf 365} 721 (1993).
\bibitem{Asbury03} 
Asbury C L, Fehr A N and Block S M  {\it Science} {\bf 302} 2130 (2003).
\bibitem{Kaseda03}
Kaseda K, Higuchi H and Hirose K  {\it Nat. Cell Biol.} {\bf 5} 1079 (2003).
\bibitem{Yildiz04}
Yildiz A, Tomishige M, Vale R D and Selvin P R  {\it Science} {\bf 303} 676 (2003).
\bibitem{Block90}
Block S M, Goldstein L S 	and Schnapp B J  {\it Nature} {\bf 348} 348 (1990).
\bibitem{Kutys10}
Kutys M L, Fricks J and Hancock W O  {\it PLoS Comp. Biol.} {\bf 6} e1000980 (2010).
\bibitem{Kramers40}
Kramers H A  {\it Physica} {\bf 7} 284 (1940).
\bibitem{Hanggi91}
H\"anggi P, Talkner P and Borkovec M  {\it Rev. Mod. Phys.} {\bf 62} 251 (1990).
\bibitem{Keller00}
Keller D and Bustamante C.  {\it Biophys. J.} {\bf 78} 541 (2000).
\bibitem{Cai07}
Cai D, Verhey K J and Meyh\"ofer E  {\it Biophys. J.} {\bf 92} 4137 (2007).
\bibitem{Hill04}
Hill D B, Plaza M J, Bonin K and Holzwarth G  {\it Eur. Biophys. J.} {\bf 33} 623 (2004). 
\bibitem{Ross16}
Ross J L  {\it Biophys. J.} {\bf 111} 909 (2016).
\bibitem{Weiss13}
Weiss M  {\it Phys. Rev. E} {\bf 88} 010101 (2013).
\bibitem{Hofling13}
Hofling F and Franosch T  {\it Rep. Prog. Phys.} {\bf 76} 046602 (2013). 
\bibitem{Gagliano10}
Gagliano J, Walb M, Blaker B, Macosko J C and Holzwarth G  {\it Eur. Biophys. J.} {\bf 39} 801 (2010). 
\bibitem{Sozanski15}
Soza\'nski K, Ruhnow F., Wi\'sniewska A., Tabaka M., Diez S. and Holyst R.  {\it Phys. Rev. Lett.} {\bf 115} 218102 (2015).
\bibitem{Goychuk12}
Goychuk I  {\it Adv. Chem. Phys.} {\bf 150} 187 (2012). 
\bibitem{Goychuk09}
Goychuk I  {\it Phys. Rev. E} {\bf 80} 046125 (2009).
\bibitem{Lebowitz99}
Lebowitz J L and Spohn H  {\it J. Stat. Phys.} {\bf 95} 333 (1999).
\bibitem{Barato15}
Barato A C and Seifert U  {\it Phys. Rev. Lett.} {\bf 114} 158101 (2015).
\bibitem{Fisher99}
Fisher M E and Kolomeisky A B  {\it Proc. Natl. Acad. Sci. USA} {\bf 96} 6597 (1999).
\bibitem{Fisher99b}
Fisher M E and Kolomeisky A B, {\it Physica A} {\bf 274}, 241 (1999).
\bibitem{Kolomeisky00}
Kolomeisky A B and Fisher M E {\it J. Chem. Phys.} {\bf 113} 10867 (2000).
\bibitem{Goychuk14a}
Goychuk I, Kharchenko V O and Metzler R  {\it PLOS One} {\bf 9} e91700 (2014).
\bibitem{Goychuk14b}
Goychuk I, Kharchenko V O and Metzler R {\it Phys. Chem. Chem. Phys.} {\bf 16} 16524 (2014).
\bibitem{Kolomeisky13}
Kolomeisky A B  {\it J. Phys.: Condens. Matter} {\bf 25} 463101 (2013).
\bibitem{Zwanzig01}
Zwanzig R  {\it Nonequilibrium Statistical Mechanics} Oxford University Press (2001).
\bibitem{Gillespie77}
Gillespie D T  {\it Phys. Lett.} {\bf 64A} 22 (1977).
\bibitem{Esposito08}
Esposito M and Lindenberg K  {\it Phys. Rev. E} {\bf 77} 051119 (2008).
\bibitem{Maes09}
Maes C, Netocny K and Wynants B  {\it J. Phys. A} {\bf 42} 365002 (2009).

\bibitem{Weber10}
Weber S C, Spakowitz A J and Theriot J A  {\it Phys. Rev. Lett.} {\bf 104} 238102 (2010).
\bibitem{Fakhri14}
Fakhri N {\it et al.}  {\it Science} {\bf 344} 1031 (2014).
\bibitem{Touchette09}
Touchette H  {\it Phys. Rep.} {\bf 478} 1 (2009).
\bibitem{Kolomeisky07}
Kolomeisky A B and Fisher M E  {\it Annu. Rev. Phys. Chem.} {\bf 58} 675 (2007).
\bibitem{Moffitt10}
Moffitt J R, Chemla Y R and Bustamante C  {\it Proc. Natl. Acad. Sci. USA} {\bf 107} 15739 (2010).
\bibitem{Moffitt14}
Moffitt J R and Bustamante C  {\it FEBS J.} {\bf 281} 498 (2014).
\bibitem{Pietzonka17}
Pietzonka P, Ritort F and Seifert U  {\it Phys. Rev. E} {\bf 96} 012101 (2017).
\bibitem{Proesmans17}
Proesmans K and Van den Broeck C  {\it Europhys. Lett.} {\bf 119} 20001 (2017). 





\end{thebibliography}
\end{document}